\documentclass[aps,prb,twocolumn,showpacs,groupedaddress,10pt]{revtex4-1}
\usepackage{natbib}
\usepackage{amsmath}
\usepackage{amssymb}
\usepackage{amsthm}
\usepackage[pdftex]{graphicx}
\usepackage{epstopdf}
\usepackage{bm}
\usepackage{hyperref}
\usepackage{epsfig}
\usepackage{dcolumn}
\usepackage{url}
\usepackage{color}

\usepackage{amsthm}%\usepackage{epstopdf}%\usepackage{subfigure}
\usepackage{bm}
\usepackage{epsfig}
\usepackage{dcolumn}

\makeatother

\begin{document}

\title{Signatures of the topological spin of Josephson vortices in topological superconductors}

\author{Daniel Ariad and Eytan Grosfeld}

\affiliation{Department of Physics, Ben-Gurion University of the Negev, Be'er-Sheva 84105, Israel}

\begin{abstract}
We consider a modified setup for measuring the Aharonov-Casher phase which consists of a Josephson vortex trapped in an annular topological superconducting junction. The junction encloses both electric charge and magnetic flux. We discover a deviation from the Aharonov-Casher prediction whose origin we identify in an additive universal topological phase that remarkably depends only on the parity of the number of vortices enclosed by the junction. We show that this phase is $\pm 2\pi$ times the topological spin of the Josephson vortex and is proportional to the Chern number. The presence of this phase can be measured through its effect on the junction's voltage characteristics, thus revealing the topological properties of the Josephson vortex and the superconducting state. 
\end{abstract}

\date{\today}

\maketitle

One of the exciting aspects of topological order is the anyonic excitations it supports, which admit fractional charge and exotic quantum statistics. Several fundamental types of anyons can be realized as vortex defects in topological superconductors, generating intensive interest in their properties \cite{Mourik2012,Das2012,Goldhaber2012}. However, detecting the anyonic properties of these vortices is an ongoing challenge. It has been proposed \cite{GrosfeldStern2011} that Josephson vortices retain the anyonic properties of bulk vortices and thus could be viable candidates for the interference experiments required to unequivocally measure their statistics. However, determining the anyon class of Josephson vortices requires finding the value of their universal exchange phase, which has not yet been reported. This exchange phase is of particular interest as it was argued that it could be used to supplement the set of quantum gates generated by the Josephson vortices to form a universal set \cite{bonderson2016twisted, Karzig2016}. 

\begin{figure}[b]
	\centering
	\includegraphics[width=0.48\textwidth]{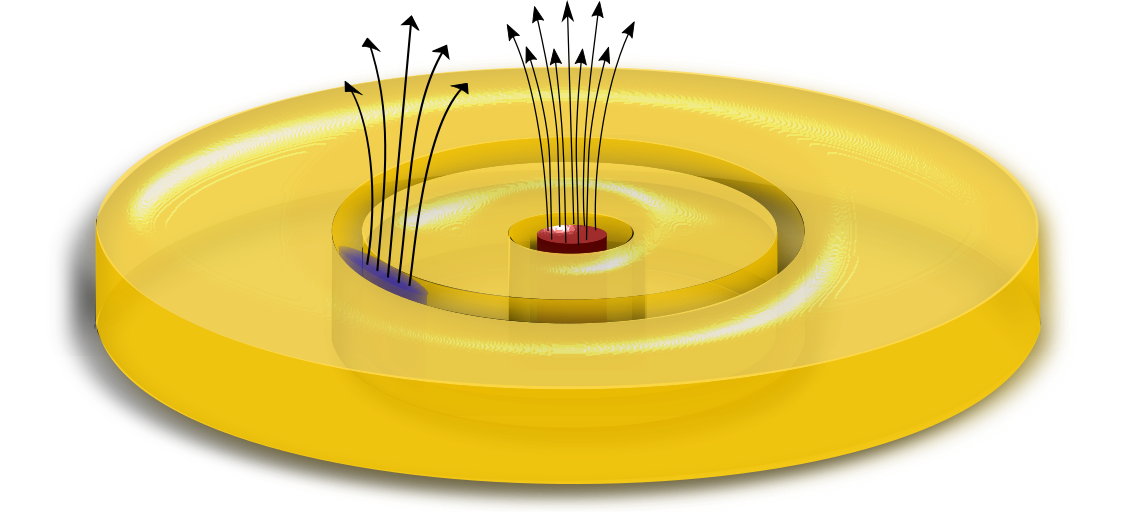}
	\caption{An annular topological Josephson junction trapping a single soliton. The soliton is depicted in blue. Counterpropagating Majorana edge states are nucleated in the junction. A charge $Q$ and phase $\Phi$ are induced externally within the central region (red).}
	\label{fig:setup}
\end{figure}

In this Rapid Communication, we report a method to calculate the universal exchange phase for Josephson vortices and propose a proof-of-principle experiment by which to measure it. 
We derive an effective quantum Hamiltonian for a Josephson vortex in a topological Josephson junction [TJJ; see Eq.~(\ref{eq:soliton-hamiltonian})], unveiling the role of the low-lying Majorana edge states in the soliton dynamics. For the case of a soliton going around an annular Josephson junction \cite{HermonSternBenJacob,Wees} (see Fig.~\ref{fig:setup}), the soliton accumulates a universal phase related to the exchange phase of Ising anyons. This phase can be exploited to induce a persistent motion of the vortex around the junction, triggered by the nucleation of an additional vortex in the region enclosed by the junction (i.e., by changing the magnetic flux $\Phi$ through the central hole). This induced motion drives the Josephson junction into its finite voltage state \cite{Wallraff2003}, revealing the presence of the phase. 

Our results therefore uncover a significant difference between nontopological Josephson junctions and TJJs. For the former, an externally induced charge $Q$ can drive the Josephson vortex into a persistent motion \cite{HermonSternBenJacob} through the Aharonov-Casher effect \cite{AharonovCasher1984,ReznikAharonov,Elion1993}. This system is analogous to an Aharonov-Bohm ring for electrons. However, the Josephson vortex remains unaffected by other vortices in the system. In contrast, for TJJs, the persistent motion of the Josephson vortex can be controlled with, instead of one knob, two: (i) continuously using the induced charge $Q$ in the region enclosed by the junction and (ii) using the enclosed flux which nucleates vortices inside the path of the vortex, hence changing their parity. In units of electron charge, the nucleation of an extra vortex within the central region is equivalent to an $e/4$ (where $e$ is the electronic charge) shift in the enclosed charge $Q$.

The dynamics of a TJJ is governed by a modified sine-Gordon Hamiltonian, where the regular bosonic degrees of freedom couple with the low-lying Majorana fermions. In particular, properties of phase solitons (Josephson vortices) through the junction are modified so that each soliton carries a Majorana zero mode \cite{FuKane2008,Grosfeld2011,GrosfeldStern2011,Tsvelik2012,Hou2012}. While experiments to probe the presence of this Majorana mode have been proposed \cite{Grosfeld2011,GrosfeldStern2011,Clarke2010}, little attention has been given to the universal properties of the host soliton itself. 

We start by discussing the fundamental mechanism behind the topological spin of a Josephson vortex. We then derive explicitly an effective Hamiltonian for the Josephson vortex and demonstrate how the topological spin plays a role in its dynamics. Next, we calculate the Berry connection governing the phase that the Josephson vortex accumulates. Finally, we propose a setup for measuring this phase.

\emph{Topological spin of the Josephson vortex.} We start by identifying the origin of the topological spin of the Josephson vortex. TJJs \cite{FuKane2008,Schnyder2008} differ from their nontopological counterparts by the presence of a pair of one-dimensional counter-propagating Majorana states present at the junction, with a Hamiltonian $H_\psi=H+\bar{H}$ ($H$ describes the external edge, and $\bar{H}$ the internal one):
\begin{eqnarray}
\nonumber H &=& i\frac{v}{2}\int dx\,\psi(x)\partial_{x}\psi(x), \\
\bar{H} &=& -i\frac{v}{2}\int dx\,\bar{\psi}(x)\partial_{x}\bar{\psi}(x).\label{eq:edge}
\end{eqnarray}
Here $x\in[0,L]$ is the coordinate of the edge, and $v$
is the neutral edge velocity. The fields obey anticommutation relations of the form
$
\{\psi(x),\psi(x')\}=\{\bar{\psi}(x),\bar{\psi}(x')\}=\delta(x-x'),
$
and $\{\psi(x),\bar{\psi}(x')\}=0$. We perform the following mode expansion, 
\begin{eqnarray} \nonumber
\nonumber \psi(x) & = & \sqrt{\frac{1}{L}}\sum_{n}e^{-2\pi inx/L}\psi_{n},\\
\bar{\psi}(x) & = & \sqrt{\frac{1}{L}}\sum_{n}e^{2\pi inx/L}\bar{\psi}_{n}.\label{mode-expansion}
\end{eqnarray}
The modes $\psi_{n}$ satisfy $\{\psi_{n},\psi_{n'}\}=\delta_{n+n',0}$
(with similar notation for the opposite chirality). Note that in particular
this implies $\psi_{0}^{2}=1/2$ (for either chirality). Plugging
this into the Hamiltonian, we get
\begin{eqnarray}
\nonumber H & = & \frac{2\pi v}{L}\left[\frac{1}{2}\sum_{n}n\psi_{-n}\psi_{n}\right]\equiv\frac{2\pi v}{L}\mathcal{L},\\
\bar{H} & = & \frac{2\pi v}{L}\left[\frac{1}{2}\sum_{n}n\bar{\psi}_{-n}\bar{\psi}_{n}\right]\equiv\frac{2\pi v}{L}\bar{\mathcal{L}}.
\end{eqnarray}
We now explore the properties of $\mathcal{L}$ and $\bar{\mathcal{L}}$, the
dimensionless momentum operators. Using
Eq.~(\ref{mode-expansion}), periodic boundary conditions on the Majorana
field imply $n\in\mathbb{Z}$, while antiperiodic boundary conditions
imply $n\in\mathbb{Z}+\frac{1}{2}$.

We examine the change in momentum when the boundary conditions are exchanged between periodic and antiperiodic for a closed circular Josephson junction, in the absence of tunneling. We write
$\mathcal{L}$ and $\bar{\mathcal{L}}$ as 
\begin{eqnarray}
\nonumber \mathcal{L} &=& \sum_{n>0}n\psi_{-n}\psi_{n}-\frac{1}{2}\sum_{n>0}n \equiv  \sum_{n>0}n\psi_{-n}\psi_{n}+\mathcal{L}_{0}(N_v),\\
\nonumber \bar{\mathcal{L}} &=& \sum_{n>0}n\bar{\psi}_{-n}\bar{\psi}_{n}-\frac{1}{2}\sum_{n>0}n \equiv  \sum_{n>0}n\bar{\psi}_{-n}\bar{\psi}_{n}+\bar{\mathcal{L}}_{0}(\bar{N}_v),\\
\end{eqnarray}
where $\mathcal{L}_{0}$ ($\bar{\mathcal{L}}_{0}$) is the ground-state contribution and $N_{v}$ ($\bar{N}_{v}$) denotes the number of vortices enclosed
by the external (internal) edge. Specifically, when there is an odd number of vortices enclosed by the edge, $n\in\mathbb{Z}$; otherwise
$n\in\mathbb{Z}+1/2$. We now calculate the difference in
the ground-state contribution in the presence of a Josephson vortex within the
junction, i.e., $N_v=1$ and $\bar{N}_v=0$. We  employ a regularizing function $F(x)$
such that $F'(x)=\partial_{x}F(x)$ decays to zero faster than $1/x^{2}$
when $x\to\infty$ and $F'(0)=1$. We calculate the regularized sum \cite{SternGrosfeldUnpublished}
\begin{eqnarray}
\nonumber\Delta\mathcal{L}_{0}&=&\mathcal{L}_{0}(1)-\bar{\mathcal{L}}_{0}(0)\\
&=&-\frac{1}{2}\sum_{n=1}^{\infty}\left[nF'(\alpha n)-(n-\frac{1}{2})F'(\alpha(n-\frac{1}{2}))\right].
\end{eqnarray}
By taking the limit $\alpha\to 0$ we now get
\begin{eqnarray}\label{eq:top-spin}
\Delta\mathcal{L}_{0} & = & -\frac{1}{2}\partial_{\alpha}\sum_{n=1}^{\infty}\left[F(\alpha n)-F(\alpha(n-\frac{1}{2}))\right]\nonumber \\
 & = & -\frac{1}{2}\partial_{\alpha}\sum_{n=1}^{\infty}\left[\frac{\alpha}{2}F'(\alpha n)-\left(\frac{\alpha}{2}\right)^{2}\frac{1}{2}F''(\alpha n)\right]\nonumber \\
 & = & -\frac{1}{2}\partial_{\alpha}\int_{\alpha/2}^{\infty}d(\alpha n)\left[\frac{1}{2}F'(\alpha n)-\frac{\alpha}{8}F''(\alpha n)\right]\nonumber \\
 & = & \frac{1}{16}\left[F'(0)+F'(\infty)\right]=\frac{1}{16}.
\end{eqnarray}
This result gives the value of the topological spin of the vortex, which is related to the dimension of the spin operator of the Ising conformal field theory (see, e.g., \cite{francesco1996conformal}). In the following we explore how this quantized momentum shift can affect the dynamics of the soliton in the presence of tunneling between the two Majorana edge states.

\emph{Effective Hamiltonian for the Josephson vortex.} We now proceed to show that the effective description of a Josephson vortex contains explicitly the topological spin discussed above. We turn on the electron tunneling across the junction, leading to a Josephson term and a Majorana tunneling term. 

The Josephson term is encapsulated in $H_\varphi$, which governs the dynamics of the relative phase degree of freedom $\varphi$ across the junction \cite{Tinkham} 
\begin{eqnarray}
H_\varphi = \frac{\hbar\bar{c}}{g^{2}}\int dx \left\{ \frac{1}{2\bar{c}^{2}}\dot{\varphi}^2+\frac{1}{2}{\varphi'}^{2}+\frac{1}{\lambda^{2}}\left[1-\cos\varphi\right]
\right\},
\end{eqnarray}
where $\dot{\varphi}\equiv(g^2\bar{c}/\hbar)\Pi$, with $\Pi$ being the canonical momentum, $\lambda$ is the Josephson penetration length, $\bar{c}$ is the renormalized velocity of light, and $g$ is a dimensionless constant which depends on the parameters of the junction \cite{HermonSternBenJacob}.

The Majorana tunneling term is first order in the electron tunneling and takes the form
\begin{equation}
H_{\text{tun}}=i\int dx\, W(x)\psi(x)\bar{\psi}(x),
\end{equation}
where $W(x)=m\cos\left[\varphi(x)/2\right]$ is the Majorana mass term \cite{FuKane2008,GrosfeldStern2011}. 

The full Hamiltonian for the TJJ, $H_{\text{TJJ}}=H_\varphi+H_\psi+H_{\text{tun}}$ \cite{GrosfeldStern2011}, is an extension of the supersymmetric sine-Gordon model for general values of $m$ \cite{Tsvelik2012}. The bosonic degrees of freedom couple with the low-lying Majorana fermions, which we now turn to solve in the presence of a single soliton.
 
We consider the solution for a classical soliton in the nonrelativistic limit which for a short and long Josephson junctions takes the approximate forms\cite{HermonShnirmanBenJacob1995}
\begin{equation}\begin{array}{ll}
\varphi_{s}(x,q(t))\simeq 2\pi\left(\frac{x-q(t)}{L}\right),& \lambda\gg L,\\
\varphi_{s}(x,q(t))\simeq 4\arctan\exp\left(\frac{x-q(t)}{\lambda}\right),& \lambda\ll L,
\end{array}\end{equation}
respectively, with a center-of-mass coordinate at $q(t)$. We plug the solution into the Euclidean action derived from the Hamiltonian $H_\varphi$ to get the energy associated with the soliton center of mass coordinate \cite{kato1996macroscopic}, $\frac{1}{2}m_{s}\dot{q}^{2}+E_{0}$,
where we defined the soliton mass $m_s$ [$m_s=(2\pi)^2\hbar/g^2\bar{c}L$ for $\lambda\gg L$ and $m_s=8\hbar/g^2\bar{c}\lambda$ for $\lambda\ll L$] and the soliton rest energy \cite{HermonSternBenJacob}. We now proceed to the Majorana sector, $H_{\psi}=\int dx\,\Psi^{T}H_0\Psi$, with $\Psi=\left(\begin{array}{cc}
\psi & \bar{\psi}\end{array}\right)^{T}$ and
\begin{eqnarray}
	H_0=\frac{1}{2}\left[\begin{array}{cc}
iv\partial_{x} & i W(x,q(t))\\
-i W(x,q(t)) & -iv\partial_{x}
\end{array}\right],\quad
\end{eqnarray}
where  $W(x,q(t))=m\cos[\varphi_s(x,q(t))/2]$. The equations simplify considerably by taking a Galilean boost to the moving frame,
\begin{eqnarray}
\nonumber & x' =  x-q(t),\quad t' =  t, \\
\nonumber & \partial_{x}  =  \partial_{x'}, \quad \partial_{t}  =  -\dot{q}\partial_{x'}+\partial_{t'}.
\end{eqnarray}
We see that the Majorana fields couple to the center-of-mass velocity of the soliton via a vector-potential-like term that measures the total momentum carried by the two counterpropagating edge states, taking the form
\begin{eqnarray}
	\frac{i}{2}\dot{q}\int dx\left(\psi\partial_{x}\psi+\bar{\psi}
\partial_{x}\bar{\psi}\right)=\frac{2\pi}{L}\dot{q}(\mathcal{L}-\bar{\mathcal{L}}).
\end{eqnarray}
The junction Hamiltonian $H_{\text{TJJ}}$, written in the background of a single soliton, is given in terms of the soliton's center-of-mass momentum $\hat{p}$ (which we now reinstate as a quantum operator) as
\begin{eqnarray}
 \nonumber H_{s}&=&E_0+\frac{1}{2m_s}\left[\hat{p}-\frac{2\pi}{L}\left(\mathcal{L}-\bar{\mathcal{L}}\right)\right]^2\\
 & & +\frac{2\pi v}{L}(\mathcal{L}+\bar{\mathcal{L}})+i\int dx\, W(x)\psi(x)\bar\psi(x). \label{eq:soliton-hamiltonian}
\end{eqnarray}
This Hamiltonian describes the dynamics of the Josephson vortex within the junction and is our first main result. The ground-state contribution to the vector potential is given by
\begin{eqnarray}
	\frac{2\pi}{L}(\mathcal{L}_0-\bar{\mathcal{L}}_0)=(-1)^{N_v}\frac{2\pi}{L}\frac{1}{16},\label{eq:berry-phase}
\end{eqnarray}
coinciding with the one calculated previously in Eq.~(\ref{eq:top-spin}). This suggests that the topological spin of the soliton affects its dynamics and may be measurable. We next turn to show that the low-lying fermion states do not affect the universality of this phase  in the adiabatic limit by providing numerical evidence.

\begin{figure}[t]
	\centering
	\includegraphics[width=0.48\textwidth]{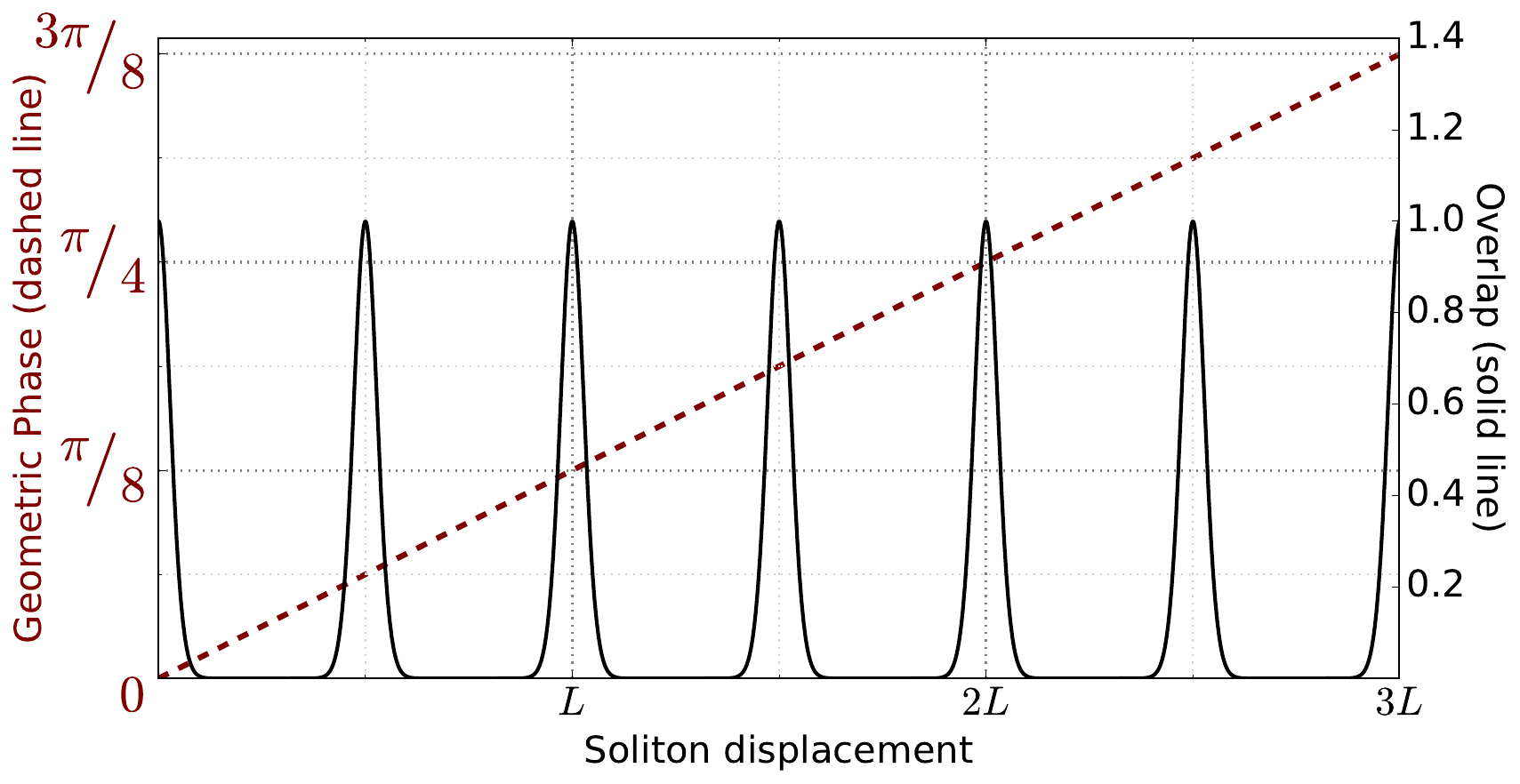}
	\caption{Numerical results for the geometric phase accumulated by the persisting Josephson vortex. The dashed brown line describes the geometric phase accumulated by each persisting soliton in the presence of a vortex within the central region. In addition, the solid black line describes the overlap norm of two counterpropagating solitons, which becomes nonzero at half cycles. At these points the geometric phase of each soliton acquires its universal values $n\pi/16$, $n\in\mathbb{Z}$. \label{fig:berry}}
\end{figure}

\emph{The Berry connection.} Due to the interactions of the Josephson vortex with the sub gap states of energies $\Delta_n$ ($n=0,1,\ldots$), the phase of the soliton is universal only when its traverse time around the junction is large compared to $\hbar/\Delta_1$. We establish this by introducing a numerical procedure for finding the Berry phase that the ground state $|\Omega_q\rangle$ accumulates as function of the position of the soliton, $q$.

We take a short Josephson junction. When the soliton goes adiabatically around the junction, the Majorana edge states depend parametrically on its position. In addition, there is a $\mathbb{Z}_2$ phase associated with the motion of the soliton: when the soliton completes a cycle, each fermionic mode enclosed by its motion acquires a minus sign. We work in momentum states and truncate the Hilbert space to retain $4N+2$ modes: $2N$ modes in the antiperiodic edge and $2N+1$ modes in the periodic edge, the latter including a Majorana zero mode $\psi_0$.  The final mode we retain is the extra Majorana zero-energy state $\psi_v$, which is localized far from the Josephson junction, either at the center of the annulus or at its outer edge, depending on the parity of the number of vortices in the central hole. In addition, we perform a gauge transformation in which the Majorana fields are single valued under $q\to q+L$ by absorbing the $\mathbb{Z}_2$ phase into the Majorana tunneling term. 

Next, we transform the Hamiltonian into a Bogoliubov form for fermions by taking appropriate superpositions of the two zero-energy Majorana fermions, $(\psi_0\pm i \psi_v)/\sqrt{2}$. The spinor is then rearranged so that particle-hole conjugation is written as $\tau_x K$ (where $\tau_x$ is the first Pauli matrix in Bogoliubov space and $K$ is complex conjugation). The Hamiltonian can then be diagonalized via
\begin{equation}
	\begin{pmatrix} H_{1} & H_{2} \\ H_{2}^\dagger & \textendash H_{1}^*\end{pmatrix}\begin{pmatrix} U & V^* \\ V & U^* \end{pmatrix}=\begin{pmatrix} E & 0 \\ 0 & -E \end{pmatrix}\begin{pmatrix}U & V^* \\ V & U^* \end{pmatrix}\!. \label{eq:bdg}
\end{equation}
The correct choice of the zero mode that is contained in the positive-energy group of $2N+1$ eigenvectors leads to a non-vanishing determinant of $U$. We can then use Eq.~(\ref{eq:bdg}) to form the BCS ground state $|\Omega_q\rangle$. Explicitly, for $q=0$, the Hamiltonian blocks are $H_{1}\!=\!\bigoplus_{k=0}^{2N}\frac{k\pi}{L}$ and 
\begin{equation}
H_{2}\!=\!\frac{m}{2}\left[0\!\oplus\!\left(\bigoplus_{k=1}^N\sigma_y\!\right)\!-\!\left(\!\frac{\sigma_y}{\sqrt{2}}\right)\!\oplus\!\left(\bigoplus_{k=1}^{N-1}\sigma_y\!\right)\oplus 0\right]\!.\nonumber
\end{equation}

 The Berry connection for $|\Omega_q\rangle$ is given by
\begin{eqnarray}
	i\langle \Omega_q| \partial_q \Omega_q\rangle=\frac{i}{4}\mbox{Tr}\left\lbrace(1+gg^\dag)^{-1}\left[g' g^\dag-g(g^\dagger)'\right]\right\rbrace, \label{eq:berry}
\end{eqnarray}
where $g=(V U^{-1})^*$\cite{read2009non}. In addition, we define the translation operator $T$ for the soliton
$\chi_q=T_q\chi_0$ with $\chi^T_q=(U^T_q, V^T_q)$.
$T_q$ is given explicitly by $T_q=Z_q P_q$, with $P_q$ generating the translation and $Z_q$ generating the $\mathbb{Z}_2$ transformation:
\begin{eqnarray}
\nonumber &&P_q=P^{(1)}\oplus P^{(2)}, \; P^{(1)}=P^{(2)*}=\bigoplus_{n=1}^{2N+1} e^{(-1)^n (1-n) i\pi q/L}, \\ 
\nonumber &&Z_q=Z^{(1)}\oplus Z^{(2)}, \; Z^{(1)}=Z^{(2)}=\bigoplus_{n=1}^{2N+1} (-1)^{\textrm{mod}(n,2)\left\lfloor\frac{q}{L}+\frac{1}{4}\right\rfloor}.
\end{eqnarray}
We diagonalize Eq.~(\ref{eq:bdg}) numerically for $q=0$, and using $T_q$ we obtain the eigenvectors for any other position of the soliton. We substitute into Eq.~(\ref{eq:berry}), performing the derivative symbolically. The result is presented in Fig.~\ref{fig:berry}  with the overlap calculated using the Onishi formula,
$|\langle \Omega_{-q}|\Omega_{q}\rangle| = \sqrt{|\det \chi^\dagger_{-q} \chi_q|}$ \cite{onishi1966generator} for two counterpersisting solitons, demonstrating that the topological spin is, in principle, an observable. We repeated the procedure taking reversed boundary conditions on the two Majorana edge states, obtaining the same phase but with an additional minus sign, which reproduces Eq.~(\ref{eq:berry-phase}) to machine precision.

\emph{Proposed setup for detecting the phase shift.} We finally consider the setup depicted in Fig.~\ref{fig:setup} where a single Josephson vortex is trapped within the junction and the voltage between the inner and outer superconducting plates is measured. The energy spectrum of the Josephson vortex can be derived from Eq.~(\ref{eq:soliton-hamiltonian}), and in the presence of an externally induced Aharonov-Casher charge $Q$ within the central region, is given by
\begin{eqnarray}
	E_s=E_c\left[\frac{Q}{2e}+\left(\frac{n_f}{4}+\frac{n_v}{16}\right)-N_b\right]^2,
\end{eqnarray}
where $E_c$ is the charging energy for the junction, $n_f=(-1)^{N_f}$ is the fermion parity within the enclosed path of the Josephson vortex ($N_f$ is the fermion number),  $n_v=(-1)^{N_v}$ is the parity of the number of vortices within the same region, and $N_b\in\mathbb{Z}$ is the relative number of Cooper pairs between the two superconducting plates. In the low energy sector there is an emergent dependence between $n_f$ and $n_v$: If $n_v=1$, then $n_f=1$, but if $n_v=-1$, then $n_f$ is free \cite{GrosfeldStern2011}.

\begin{figure}[t]
	\centering
	\includegraphics[width=0.48\textwidth]{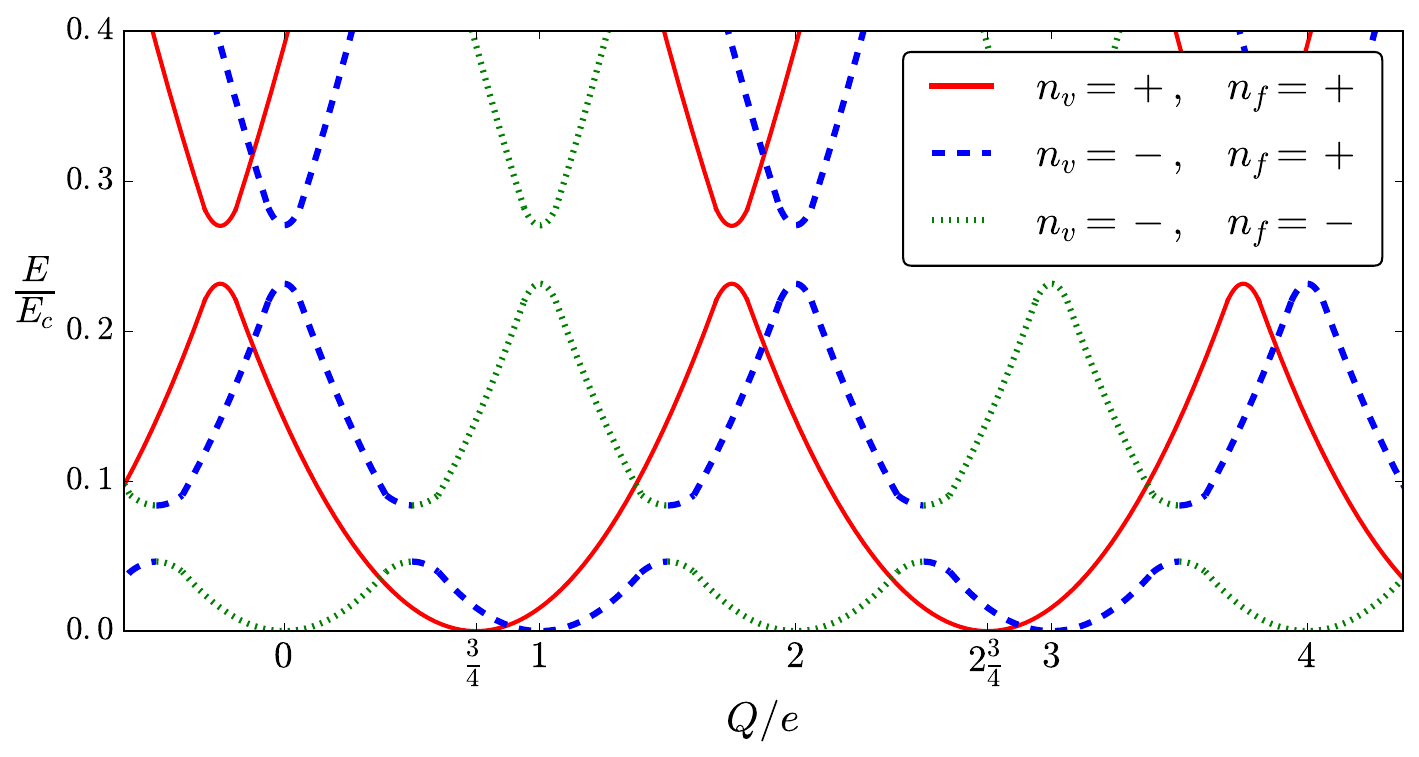}
	\caption{Energy spectrum for the Josephson vortex. Solid red lines describe the energy of the Josephson vortex in the presence of an even number of vortices enclosed within its path. Dashed blue and dotted green lines describe the case with an odd number of vortices (for even and odd fermion parities, respectively). The velocity of the persisting soliton is proportional to the gradient of the energy, $v_s\propto\partial_Q E$. Fermion parity changing effects open a gap between the green and blue lines, disorder opens a gap between lines of the same color.}
		\label{fig:graph}
	\end{figure}

Assume we start from the case that there are no vortices within the central hole in the annulus (Fig.~\ref{fig:setup}), i.e., $n_v=1$ and $n_f=1$. The junction can be tuned into the zero voltage state by shifting the induced Aharonov-Casher charge $Q$. The Josephson vortex accordingly acquires a vanishing velocity. Next, we add an extra vortex within the central region of the sample, shifting the value of $n_v$ to $-1$. The Josephson vortex acquires a phase shift which is equivalent to a $\pm e/4$ shift in the induced Aharonov-Casher charge (see Fig.~\ref{fig:graph}). It then performs a persistent motion, and the junction is driven into its finite voltage state. This dependence of the voltage characteristics of the junction on the number of vortices enclosed within the junction is our second main result.

One possible realization of the system is a topological insulator with an $s$-wave superconductor deposited on its surface, forming a Josephson junction shaped as in Fig.~\ref{fig:setup}. The dynamics of the soliton will be largely determined by the $s$-wave superconducting layer, while a Majorana zero mode will be trapped by the soliton on the surface state of the topological insulator. Furthermore, the charge on the central island will be varied by means of a capacitive gate\cite{Elion1993}.

\emph{Discussion.} Our central result is the identification of a relative $\pi/4$ phase associated with a Josephson vortex in a topological Josephson junction encircling an odd versus even number of vortices. It is useful to compare this result with the full conformal case which describes the physics with a vanishing Majorana mass, $m=0$. Then, vortex exchange is captured by a standard fusion rule from conformal field theory (see, e.g., \cite{francesco1996conformal}), $\sigma(z)\sigma(0)\sim z^{-1/8}\left[I+z^{1/2}\psi(z)\right]$, where $I$ is the identity field and $\psi$ and $\sigma$ are fields of dimensions $1/2$ and $1/16$ respectively. By identifying the field $\sigma(z)$ as the vortex and $z=x+i y$ as its coordinate, this equation reproduces the presence of a $-\pi/4$ phase shift for a rotation of one vortex around another, $z\to e^{2\pi i}z$. For the case of an odd fermionic number, a $3\pi/4$ phase shift would ensue. Instead, in our case, the nonzero Majorana mass term protects the anyon properties decided by the bulk topological quantum field theory, which is a manifestation of Ocneanu rigidity.\cite{Kitaev20062}

Finally, we address the context of this work from experimental and theoretical perspectives. Trapping a single Josephson vortex within an annular Josephson junction has been experimentally achieved \cite{Wallraff2003, Fedorov2014}. It was demonstrated that the Josephson vortex is able to tunnel through a barrier, revealing its quantum nature \cite{Wallraff2003}. Interference experiments of Josephson vortices have been reported \cite{Elion1993}. Recently, Josephson vortices were directly observed with scanning tunneling spectroscopy, and their local density of states was deduced \cite{Roditchev2015}. More specifically, in the context of topological superconductors, quasiparticle poisoning may affect observables that are sensitive to fermion parity-changing effects. However, the $e/4$ shift discussed here remains immune to a shift by $e$, and hence so is the residual motion of the soliton generated by it. Possible realizations of annular topological Josephson junctions were discussed in \cite{GrosfeldStern2011} using semiconductor heterostructures or $p$-wave superconductors (see, e.g., \cite{Maeno2012}). Solitons in other scenarios involving $p$-wave superconductors and two-band superconductors were discussed in \cite{Kaneyasu2010,Vakaryuk2012}. Other papers touching on the Aharonov-Casher effect in topological superconductors include \cite{hassler2010anyonic,bonderson2011topological}. The effective action of bulk Abrikosov vortices was considered in \cite{ariad2015effective}. 

This research is supported by the Israel Science Foundation (Grants  No.~401/12 and  No.~1626/16), the European Union's Seventh Framework Programme (FP7/2007-2013) under Grant No.~303742, and the Binational Science Foundation through Grant No.~2014345.

\bibliographystyle{apsrev}
\bibliography{soliton}

\end{document}